\documentclass[lettersize,journal]{IEEEtran}
\usepackage{amsmath,amsfonts}
\usepackage{algorithmic}
\usepackage{algorithm}
\usepackage{array}
\usepackage[caption=false,font=normalsize,labelfont=sf,textfont=sf]{subfig}
\usepackage{textcomp}
\usepackage{stfloats}
\usepackage{url}
\usepackage{verbatim}
\usepackage{graphicx}
\usepackage{cite}

\usepackage{stfloats}
\usepackage{lineno}
\usepackage{multirow}
\usepackage{graphicx}

\usepackage{multirow}
\usepackage{graphicx}

\usepackage{amsmath,amssymb,graphicx,epsf,color,url,booktabs,tabularx}
\usepackage{psfig,float}
\usepackage[justification=centering]{caption}
\usepackage{amsmath,amssymb}
\usepackage{multirow}

\usepackage[numbers,sort&compress]{natbib}

\usepackage{amsfonts}
\usepackage{algorithmic}
\usepackage{textcomp}
\usepackage{xcolor}
\usepackage{color}

\makeatletter
\renewcommand{\maketag@@@}[1]{\hbox{\m@th\normalsize\normalfont#1}}%
\makeatother

\begin{document}

\title{Joint Waveform Design for MIMO-OFDM DFRC Systems}

\DeclareRobustCommand*{\IEEEauthorrefmark}[1]{%
    \raisebox{0pt}[0pt][0pt]{\textsuperscript{\footnotesize\ensuremath{#1}}}}
    
%
%
%
%
\author{Tianchen Liu,Yifei Liu, Liang Wu, {\it Senior Member, IEEE}, Bo An, Zaichen Zhang, {\it Senior Member, IEEE}, Jian Dang, {\it Senior Member, IEEE} and Jiangzhou Wang, {\it Fellow, IEEE}

\thanks{T. Liu, Y. Liu, L. Wu, B. An, Z. Zhang and J. Dang are with National Mobile Communications Research Laboratory, Southeast University, Nanjing 210096, China (e-mail: \{220210764, yifei\_liu, wuliang, anbo, zczhang, dangjian\}@seu.edu.cn).}
\thanks{J. Wang is with School of Engineering, University of Kent, Canterbury, CT2 7NT United Kingdom (e-mail: j.z.wang@kent.ac.uk).}
\thanks{Tianchen Liu and Yifei Liu equally contribute to this work.}
}



\maketitle

 \begin{abstract}
  Dual-functional radar-communication (DFRC) has attracted considerable attention. This paper considers the frequency-selective multipath fading environment and proposes DFRC waveform design schemes based on multiple-input and multiple-output (MIMO) and orthogonal frequency division multiplexing (OFDM) techniques. In the proposed waveform design schemes, the Cramer-Rao bound (CRB) of the radar system, the inter-stream interference (ISI) and the achievable rate of the communication system, are respectively considered as the performance metrics. In this paper, we focus on the performance trade-off between the radar system and the communication system, and the optimization problems are formulated. In the ISI minimization based waveform design scheme, the optimization problem is convex and can be easily solved. In the achievable rate maximization based waveform design scheme, we propose a water-filling (WF) and sequential quadratic programming (SQP) based algorithm to derive the covariance matrix and the precoding matrix. Simulation results validate the proposed DFRC waveform designs and show that the achievable rate maximization based scheme has a better performance than the ISI minimization based scheme.
 \end{abstract}

 \begin{IEEEkeywords}
Dual-functional radar-communication, multiple-input and multiple-output, orthogonal frequency  division multiplexing, multipath fading.
\end{IEEEkeywords}

\section{Introduction}
\label{s1}

Due to the shortage of radio spectrum, radar and communication systems are expected to use the same frequency band \cite{ZChen,PZhang,LWang}. In addition, both the hardware architecture and the system composition of radar system and communication system are similar \cite{BPaul}. Therefore, the concept of joint radar and communication (JRC) has been proposed with the aim of miniaturization and commercialization. It is expected that JRC technology will promote huge potential applications. For example, recently, drones have been widely used in transportation and geographic explorations. However, drones with aggressive purposes pose a threat to social security \cite{XShi}\cite{IGuvenc}. Thus, drone surveillance radars are usually deployed. By employing JRC technology, drone surveillance radars not only have the advantage of quick deployment, but also interact with each other by establishing communication links to improve the target-detection accuracy. Moreover, to realize automatic driving, the transmission delay should be less than 10ms, and the position accuracy should be better than 1 centimeter \cite{HWym}. JRC technology is expected to provide stable and precise sensing to guarantee the safety of autonomous driving, and the communication link is established at the same time to reduce the latency \cite{PKumari}. Furthermore, JRC technology can reduce the number of antennas, and alleviate the problem of electromagnetic interference or compatibility \cite{ZFeng}. Therefore, JRC technology will play an important role in future mobile communication and radar detection systems.

The waveform design, which is one of the most challenging researches in the JRC system, is mainly classified into two categories. The first category is that the radar system and the communication system co-exist \cite{TCCN1}\cite{YHe}, where one system treats the other system as interference \cite{FLiu1}. Therefore, multiplexing techniques including space division multiplexing (SDM) \cite{AHass1,AHass2,AHass3}, time division multiplexing (TDM) \cite{HTaka,LHan1,LHan2}, frequency division multiplexing (FDM) \cite{AMishra}\cite{LReichardt}, and code division multiplexing (CDM) \cite{HTakase}\cite{MJamil} have been applied to mitigate the interference in the co-existing system. However, in these multiplexing schemes, the co-existing waveform has a low utilization because the radar system and the communication system cannot employ the same resources, and two different kinds of waveforms are usually used respectively for radar and communication. The second category is that the radar system and the communication system are co-designed, which aims to develop a single waveform that can perform radar detection and communication functions simultaneously \cite{FLiu2}\cite{Renz}\cite{Huah}, and this waveform is called dual-functional radar-communication (DFRC) waveform. Compared with the co-existing waveform, the DFRC waveform has a higher degree of integration, thus it is more attractive to both academia and industry \cite{JAndrew}.

The DFRC waveform design includes the radar-centric design, the communication-centric design, and the joint waveform design. The main idea of the radar-centric design is to employ linear frequency modulation (LFM) signal, which is a traditional radar probing signal, as the communication information carrier. For example, amplitude shift keying (ASK) and phase shift keying (PSK) can be combined with LFM to design the DFRC waveform in \cite{MRoberton}\cite{GN}. In the communication-centric design, the traditional communication waveform is mainly employed, and the detection estimation parameters can be extracted through the echo of the transmitted communication signal \cite{DMa,CSturm,DGaglione,FLiu3}. Joint waveform design is neither based on existing radar waveform nor communication waveform, but jointly considers the performance of the DFRC system and aims to achieve a performance trade-off between the communication and the radar detection \cite{FLiu2}, \cite{FLiu3,XLiu,XWang,FLiu4}.

A great quantity of work has been devoted to the research of the joint waveform design. In \cite{FLiu2}, mutual communication user interference (MUI) was employed as the communication performance metric, and the mismatch degree between the actual beam pattern and the desired beam pattern was used to evaluate the radar performance. A low-complexity algorithm was proposed to balance the radar performance and the communication performance. In \cite{XLiu}, the cross-correlation of the beam strength at different angles was involved as a new radar performance metric and the precoding matrices of the radar system and the communication system were optimized to achieve the performance trade-off. \textsl{Wang et al.} in \cite{XWang} employed the system model and the performance matrix of \cite{FLiu2}, and used a reconfigurable intelligence surface (RIS) to achieve an additional degree of freedom (DoF). The Cramer-Rao bound (CRB) of the radar system was employed in \cite{FLiu4}, and a higher beam directivity was achieved. In \cite{TZhang} and \cite{MJiang}, the sidelobe level of the transmit beam pattern was minimized, and the peak-to-average power ratio (PAPR) of the transmit signal was used as a constraint to guarantee the power amplification efficiency of the high power amplifiers (HPAs).

All the above works adopted the single-carrier modulation, and the flat fading channel model was used. However, the frequency-selective multipath fading channel is more reasonable in the high-rate transmission. Orthogonal frequency division multiplexing (OFDM) \cite{HZhu1}\cite{HZhu2} is one of the key technologies in the fifth generation (5G) mobile communication for its robustness to the multipath fading. Besides, it has been proven to be valid for the radar detection in \cite{CSturm}. \textsl{Xu et al.} in \cite{ZYXu} presented a complete processing strategy for the measurement of range, velocity, and angle based on OFDM signals. Meanwhile, a trade-off scheme that employed dedicated subcarriers to balance the radar performance and the communication performance was also proposed. However, different subcarriers in \cite{ZYXu} used the same precoding matrix, which led to a performance degradation in the DFRC system. The beam strength radiated towards the detection target is usually greater than that towards the communication user. Therefore, a subcarrier power allocation scheme was proposed in \cite{TLiu} to satisfy the different power requirements of the radar system and the communication system. \textcolor{black}{However, \cite{TLiu} focuses on achieving the omnidirectional radiation in the radar detection phase, and it does not consider how to minimize CRB.} \textsl{Hu et al.} in \cite{XHu} extended the work of \cite{FLiu2} and the PAPR of the designed OFDM waveform was optimized to achieve a better communication performance.

In this paper, we consider a single communication user MIMO system in the frequency-selective multipath fading environment. The DFRC base station (BS) utilizes a single waveform to realize the downlink communication and the target detection simultaneously. Because the MUI minimization scheme used in \cite{FLiu2} cannot be directly used in the frequency-selective multipath fading environment, we propose an OFDM based waveform design scheme, which can effectively combat the multipath fading, to minimize the inter-stream interference (ISI) of the communication system and the CRB of the radar system. Furthermore, the achievable rate maximization based waveform design scheme and the joint water-filling and sequential quadratic programming (WF-SQP) algorithm are proposed to further improve the communication performance through the cooperation of different antennas.

The main contributions of this paper are summarized as follows:

1. We focus on the performance trade-off between the communication system and the radar system, and propose two OFDM based DFRC waveform design schemes to combat the frequency-selective multipath fading. In the first proposed DFRC waveform design scheme, we minimize the ISI of the communication system and the CRB of the radar system. The algorithms for designing the transmit covariance matrix and the precoding matrix design algorithms are proposed.

2. Different from the first DFRC waveform design scheme, the achievable rate of the communication system is employed in the second DFRC waveform design scheme. The performance trade-off is also considered. To derive the optimal transmit covariance matrix, a joint water-filling and SQP algorithm is proposed. Furthermore, the precoding matrix is derived analytically.

3. Simulation results are provided and analyzed in detail. Simulation results show that the proposed DFRC waveform design schemes can achieve satisfied communication and radar performances. The second proposed DFRC waveform design scheme achieves better radar and communication performances than the first one.

The rest of this paper is organized as follows. Section \ref{s2} presents the MIMO-OFDM based system model. In Section \ref{s3}, we propose the first OFDM based DFRC waveform design scheme with the aim of minimizing the ISI of the communication and the CRB of the radar.  In Section \ref{s4}, we maximize the achievable rate of the communication system, and propose the second waveform design scheme. Simulation results are provided in Section \ref{s5}. Finally, the conclusion is drawn in Section \ref{s6}.

\section{System model}
\label{s2}

\begin{figure}[t]
\centering
\includegraphics [width=1\linewidth]{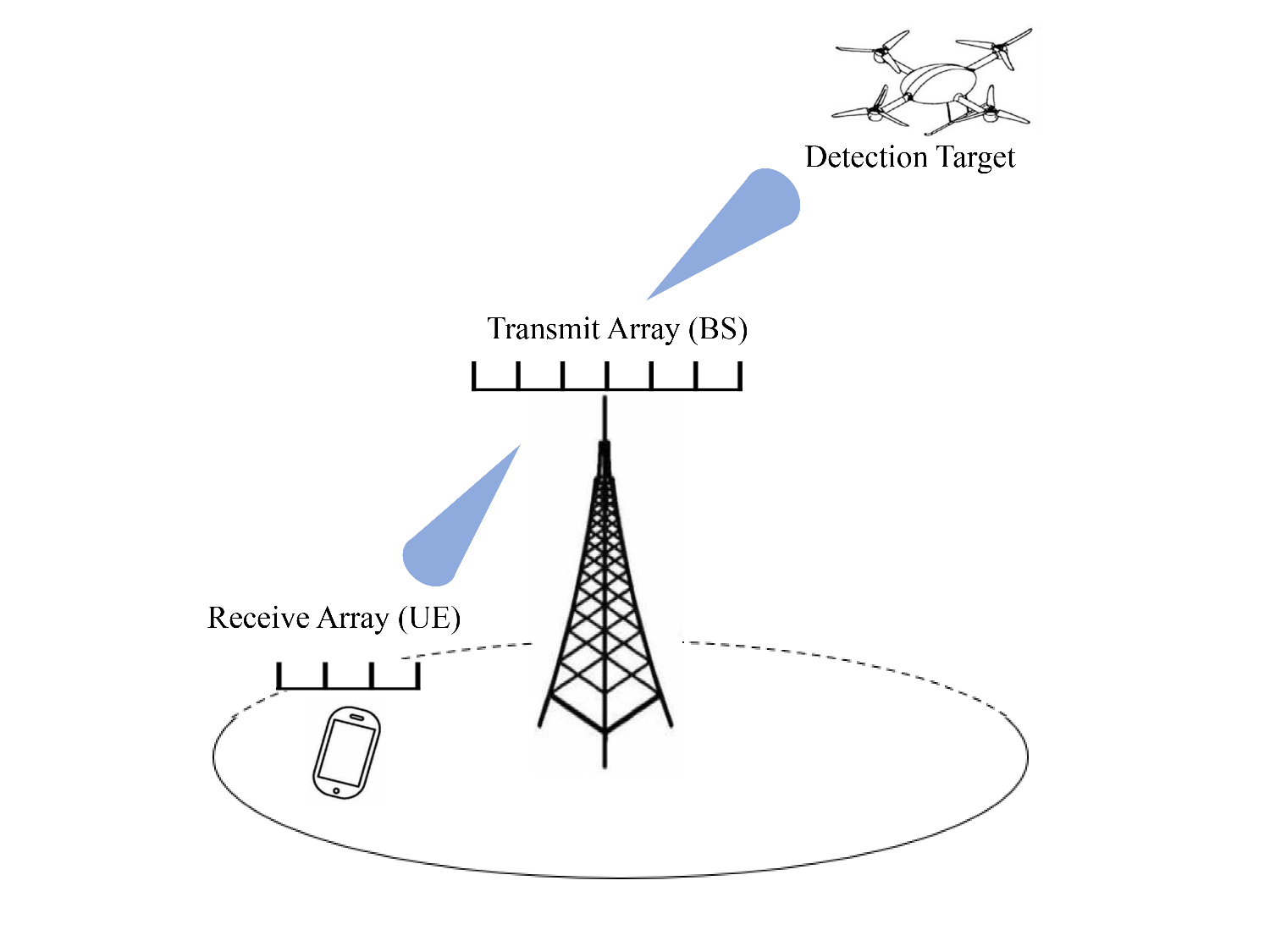}
\caption{Dual-functional radar-communication system.}
\label{Fig_sys}
\end{figure}

Consider a MIMO-OFDM DFRC system, the BS is equipped with ${{N}_{T}}$ antennas, and a uniform linear array (ULA) is employed. As shown in Fig. \ref{Fig_sys}, the BS transmits signals for the target detection and the downlink communication at the same time. It is assumed that the number of the antennas equipped at the communication user equipment (UE) is ${{N}_{R}}$, and ${{N}_{R}<{N}_{T}}$.

The transmit signal vector $\boldsymbol{x}(n)$ is expressed as
\begin{small}
\begin{equation}
\label{eqng}
\boldsymbol{x}(n) = {[{x_1}(n),{x_2}(n), \cdots ,{x_{{N_T}}}(n)]^T}, n = 1,2, \ldots ,{N-1},
\end{equation}
\end{small}
where $N$ is the size of inverse discrete Fourier transform (IDFT), $x_i(n)$ is the time-domain OFDM signal transmitted through the $i$th antenna, and it is given by
\begin{equation}
\label{eqn2}
{x_i}(n) = {1 \over N}\sum\limits_{k = 0}^{N - 1} {{X_i}(k)} {e^{j{{2\pi } \over N}nk}},
\end{equation}
and ${X_i}(k)$ is the frequency-domain signal at the $k$th subcarrier of the $i$th antenna.

The frequency-domain signal vector $\boldsymbol{x}_f(k)$ and the time-domain signal vector $\boldsymbol{x}(n)$ are expressed as
\begin{small}
\begin{subequations}
\begin{align}
\label{XX}
\boldsymbol{x}_f(k) = {\left[ {{X_1}(k),{X_2}(k), \cdots ,{X_{{N_T}}}(k)} \right]^T},&k = 0,1, \ldots ,N - 1, \\
\boldsymbol{x}_f(k) = \boldsymbol{W}(k)&\boldsymbol{s}(k),
\end{align}
\end{subequations}
\end{small}
where  $\boldsymbol{W}(k) \in {\mathbb{C}^{{N_T} \times {N_R}}}$ is the precoding matrix, and $\boldsymbol{s}(k) \in {\mathbb{C}^{{N_R}}}$ is the information vector.

The time-domain covariance matrix of the transmit signal is expressed as
\begin{subequations}
\begin{align}
&\quad\quad\quad\quad\quad\boldsymbol{R} = E\{ \boldsymbol{x}(n){\boldsymbol{x}^H}(n)\} \label{eqn4a} \\
\boldsymbol{R}(i,s) &= {1 \over {{N^2}}}E\left\{ {\sum\limits_{{k_1}=0}^{N - 1} {\sum\limits_{{k_2}=0}^{N - 1} {{X_i}({k_1})} } X_s^*({k_2}){e^{j{{2\pi } \over N}n({k_1} - {k_2})}}} \right\}, \label{eqn4b}
\end{align}
\end{subequations}
where $E\{\cdot\}$ denotes the expectation of a random variable, $(\cdot)^H$ is the conjugate transpose operator, $(\cdot)^*$ is the conjugate operator and $\boldsymbol{R}(i,s)$ is the $(i,s)$th entry of $\boldsymbol{R}$. It is supposed that each antenna employs all the subcarriers, and $E\{ {{X}_i}({k_1}){X}_s^*({k_2})\} = 0$ holds when $k_1 \neq k_2$. Therefore, Eq. (\ref{eqn4a}) can be rewritten as
\begin{equation}
\label{eqn5}
\begin{aligned}
\boldsymbol{R} &= {1 \over {{N^2}}}E\left\{ {\sum\limits_{k = 0}^{N-1} {\boldsymbol{W}(k)\boldsymbol{s}(k){\boldsymbol{s}^H}(k){\boldsymbol{W}^H}(k)} } \right\} \\
& = {1 \over {{N^2}}}\sum\limits_{k = 0}^{N-1} {\boldsymbol{W}(k){\boldsymbol{P}_S}(k){\boldsymbol{W}^H}(k)},
\end{aligned}
\end{equation}
where ${\boldsymbol{P}_S}(k) = {\rm{diag}}\left\{ {{p_{{s_1}(k)}}, \cdots, {p_{{s_{{N_R}}}(k)}}} \right\}$, ${p_{{s_i}(k)}}$ is the power of $i$th information stream at the $k$th subcarrier, and $\rm{diag}\{\cdot\}$ is the diagonal operator. The frequency-domain covariance matrix of the transmit signal at the $k$th subcarrier is defined as ${\boldsymbol{R}_f}(k) \in {\mathbb{C}^{{N_T} \times {N_T}}}$ , and it is given by
\begin{equation}
\label{eqn6}
\boldsymbol{R}_f(k) = E\{ \boldsymbol{x}_f(k){\boldsymbol{x}_f^*}(k)\}=\boldsymbol{W}(k){\boldsymbol{P}_S}(k){\boldsymbol{W}^H}(k).
\end{equation}

The relationship between the frequency-domain and the time-domain covariance matrices can be expressed as
\begin{equation}
\label{eqn7}
{1 \over {{N^2}}}\sum\limits_{k = 0}^{N-1} {{\boldsymbol{R}_f}(k)}  = \boldsymbol{R}.
\end{equation}

\subsection{MIMO-OFDM Communication Model}
The frequency-selective multipath fading channel model is employed, and the channel impulse response from the $i$th transmit antenna to the $s$th receive antenna is ${h_{s,i}}(n)$. The received signal at the $s$th receive antenna is given by
\begin{equation}
\label{eqn8}
{y_s}(n) = \sum\limits_{i = 0}^{{N_T}-1} {{h_{s,i}}(n) \otimes {x_i}(n)}  + {z_s}(n),s = 1,2, \ldots ,{N_R},
\end{equation}
where $\otimes$ denotes the convolution operation, and ${z_s}(n)$ is the noise component. Inter-subcarrier interference will increase the symbol error rate (SER) of the communication subsystem, and it will also cause inaccurate parameter estimation of the radar subsystem. It is assumed that ideal time-frequency synchronization is achieved at the receiver. In the frequency-domain, the received signal vector at the $k$th subcarrier can be expressed as
\begin{equation}
\label{eqn9}
\begin{aligned}
\boldsymbol{Y}&(k) = \boldsymbol{H}(k)\boldsymbol{x}_f(k) + \boldsymbol{Z}(k)\\
&= \boldsymbol{H}(k)\boldsymbol{W}(k)\boldsymbol{s}(k) + \boldsymbol{Z}(k),k = 0,1, \ldots ,N - 1,
\end{aligned}
\end{equation}
where ${\boldsymbol{H}}(k) \in {\mathbb{C}^{{N_R} \times {N_T}}}$ is the frequency-domain channel matrix corresponding to the $k$th subcarrier, ${\boldsymbol{Z}}(k) \in {\mathbb{C}^{N_R}}$ is the frequency-domain Guassian noise vector with covariance matrix $\sigma _{\boldsymbol{Z}}^2{{\boldsymbol{I}}_{{N_R}}}$ , and ${\boldsymbol{I}}_{{N_R}}$ is an identity matrix with rank $N_R$. It is assumed that the base station utilizes the feedback information from communication users to obtain $\boldsymbol{H}(k)$.

The vector $\boldsymbol{Y}(k)$ can be rewritten as \cite{FLiu2} 
\begin{equation}
\label{eqn10}
\boldsymbol{Y}(k) = \boldsymbol{s}(k) + \underbrace {\boldsymbol{H}(k)\boldsymbol{W}(k)\boldsymbol{s}(k) - \boldsymbol{s}(k)}_{{\rm{ISI}}} + \boldsymbol{Z}(k).
\end{equation}

The total power of ISI is
\begin{equation}
\label{eqn11}
{P_{\rm{ISI}}}   \!=\! \sum\limits_{k = 0}^{N - 1} {{{\left\| {\boldsymbol{H}(k)\boldsymbol{W}(k)\boldsymbol{s}(k) \!-\! \boldsymbol{s}(k)} \right\|}^2}}\!.
\end{equation}

Therefore, when the ISI minimization principle is employed, the achievable rate of the system is given by

\begin{footnotesize}
\begin{equation}
\label{eqn12}
{C_s} =\gamma \sum\limits_{k = 0}^{N - 1} \sum\limits_{i = 0}^{{N_R-1}} {{{\Delta f} \over B}} {{\log }_2}\left(\! {1 \!+\! {{E\{ {{{\left\| {{\boldsymbol{s}_i}(k)} \right\|}^2}} \}} \over {E\{ {{{\left\| {{\boldsymbol{H}_i}(k)\boldsymbol{W}(k)\boldsymbol{s}(k) - {\boldsymbol{s}_i}(k)} \right\|}^2}} \} + \sigma_{\boldsymbol{Z}}^2}}} \!\right),
\end{equation}
\end{footnotesize}
where $\gamma={N \over {N + {N_{\rm{CP}}}}}$, $N_{\rm{CP}}$ denotes the length of OFDM cyclic prefix, ${\Delta f}$ is the subcarrier spacing, $B$ is the bandwidth of the OFDM system, ${\boldsymbol{s}_i}(k)$ is the $i$th entry of ${\boldsymbol{s}}(k)$, and ${\boldsymbol{H}_i}(k)$ is the $i$th row of ${\boldsymbol{H}}(k)$. Note that the effect of the cyclic prefix to the data rate is ignored.

In the considered single user MIMO communication, the received signals of all antennas can be processed jointly. Therefore, the achievable rate in Eq. (\ref{eqn12}) can be increased. The maximum achievable rate of the considered single user MIMO communication system is given by \cite{SHa}

\begin{footnotesize}
\begin{equation}
\label{eqn13}
{C_t} = \gamma \cdot {\rm{max}} \bigg\{ \sum\limits_{k = 0}^{N - 1} {{{\Delta f} \over B}}{{\log }_2}\left( {\det\left( {{\boldsymbol{I}_{{N_R}}} + {{\boldsymbol{H}(k){\boldsymbol{R}_f}(k){\boldsymbol{H}^H}(k)} \over {\sigma _{\boldsymbol{Z}}^2}}} \right)} \right) \bigg \},
\end{equation}
\end{footnotesize}
where $\rm{det}(\cdot)$ denotes the determinant of a matrix.

Besides, the proposed scheme can be applied in the multiple communication user scenario, and time division multiple access (TDMA) or orthogonal frequency division multiple access (OFDMA) techniques can be used.

\subsection{MIMO Radar Model}
It is assumed that the detection target is a point target, and the echo signal received at the DFRC BS is given by
\begin{equation}
\label{eqn14}
\begin{aligned}
{\boldsymbol{y}_r}(n) &= \zeta \boldsymbol{a}(\theta ){\boldsymbol{a}^H}(\theta )\boldsymbol{x}(n) + {\boldsymbol{z}_r}(n)\\
& = \zeta \boldsymbol{A}(\theta )\boldsymbol{x}(n) + {\boldsymbol{z}_r}(n),
\end{aligned}
\end{equation}
where $\zeta$ is the reflection coefficient	and contains the information of the radar cross section (RCS) of the target, ${\boldsymbol{z}_r}(n)$ is the Gaussian noise vector \textcolor{black}{term which includes multipath interference from the same target}, $\boldsymbol{A}(\theta)=\boldsymbol{a}(\theta)\boldsymbol{a}(\theta)^H$, $\theta$ is the angle of arrival of the target, and the steering vector $\boldsymbol{a}(\theta)$ is given by \cite{FLiu4}
\begin{equation}
\label{eqn15}
\begin{aligned}
\boldsymbol{a}(\theta ) =& [{e^{ - j2\pi {{({N_T} - 1)} \over 2}\Delta \sin \theta }},{e^{ - j2\pi {{({N_T} - 3)} \over 2}\Delta \sin \theta }}, \\
&\cdots ,{e^{j2\pi {{({N_T} - 1)} \over 2}\Delta \sin \theta }}{]^T}\in {\mathbb{C}^{{N_t} \times 1}},
\end{aligned}
\end{equation}
where the center of the ULA antennas is chosen as the reference point, and $\Delta$ is the normalized antenna spacing. \textcolor{black}{It should be noted that the normalized steering vector for different subcarriers may have slight differences, which are neglected in this paper because the frequency difference between subcarriers is much less than the carrier frequency.}

CRB represents the lower bound of the variance of an unbiased estimator, and is an important indicator of the radar performance. In this paper, a smaller CRB represents a stronger stability of the radar parameter estimation. The CRB of $\theta$ can be derived as \cite{IBekk}
\begin{footnotesize}
\begin{equation}
\label{eqn16}
\begin{aligned}
&\rm{CRB}(\theta ) = {1 \over {2\rm{SNR}}} \cdot \\
&{{\rm{tr}\left( {{\boldsymbol{A}^H}(\theta )\boldsymbol{A}(\theta )\boldsymbol{R}} \right)} \over {\left( {\left( {\rm{tr}\left( {{{\boldsymbol{\dot A}}^H}(\theta )\boldsymbol{\dot A}(\theta )\boldsymbol{R}} \right)\rm{tr}\left( {{\boldsymbol{A}^H}(\theta )\boldsymbol{A}(\theta )\boldsymbol{R}} \right)} \right) - {{\left| {\rm{tr}\left( {{{\boldsymbol{\dot A}}^H}(\theta )\boldsymbol{A}(\theta )\boldsymbol{R}} \right)} \right|}^2}} \right)}},
\end{aligned}
\end{equation}
\end{footnotesize}
where $\boldsymbol{\dot A}(\theta ) = {{\partial \boldsymbol{A}(\theta )} \over {\partial \theta }}$, and $\rm{tr}(\cdot)$ is the operator of matrix trace. Substituding $\boldsymbol{A}(\theta)=\boldsymbol{a}(\theta)\boldsymbol{a}(\theta)^H$ into Eq. (\ref{eqn16}), the CRB of $\theta$ can be rewritten as \cite{IBekk}
\begin{equation}
\label{eqn17}
\rm{CRB}(\theta ) = {1 \over {2\rm{SNR}{{\left\| {\boldsymbol{\dot a}(\theta )} \right\|}^2}\boldsymbol{a}{{(\theta )}^H}\boldsymbol{R}\boldsymbol{a}(\theta )}}.
\end{equation}

According to Eq. (\ref{eqn17}), the CRB of $\theta$ relies on the covariance matrix $\boldsymbol{R}$.
\section{Joint waveform design based on CRB and ISI minimization}
\label{s3}
In this section, we first design the time-domain covariance matrix to minimize the CRB of $\theta$. After that, an optimization problem of minimizing the interference among different information streams for a given time-domain covariance matrix is formulated, and the frequency-domain covariance matrix and the corresponding precoding matrix are designed. Besides, we propose the joint waveform design scheme based on the CRB and ISI minimization to achieve the performance trade-off.
\subsection{Radar-only problem formulation}
It is assumed that the detection target is located at the angle of $\theta$. To minimize the CRB under the power constraint, the optimization problem can be formulated as follows
\begin{subequations}
\begin{align}
\rm{OP1:} \quad \mathop {\arg \min }\limits_{\boldsymbol{R}}& {1 \over {2\rm{SNR}{{\left\| {\boldsymbol{\dot a}(\theta )} \right\|}^2}\boldsymbol{a}{{(\theta )}^H}\boldsymbol{R}\boldsymbol{a}(\theta )}} \label{eqn18a}\\
&\, {\rm{s.t.}} \,\,\,\,{\rm{tr}}(\boldsymbol{R}) = {P_T}, \label{eqn18b}\\
&\quad\quad\boldsymbol{R} = {\boldsymbol{R}^H}, \label{eqn18c}
\end{align}
\end{subequations}
where ${P_T}$ is the total transmit power. The optimization problem OP1 is convex \cite{byd} and has a closed-form solution. The optimal time-domain covariance matrix is given by
\begin{equation}
\label{eqnrd}
\boldsymbol{R}_d={{{P_T}} \over {{N_T}}}{\boldsymbol{a}^*}(\theta ){\boldsymbol{a}^T}(\theta ).
\end{equation}

Although we focus on the single detection target scenario in this paper, the proposed scheme can also be extended to the multi-target scenario.

\subsection{Radar-strict problem formulation and the precoding matrix design}
Based on $\boldsymbol{R}_d$, the problem of minimizing the interference among different information streams is given by
\begin{subequations}
\begin{align}
\rm{OP2:} \quad\mathop {\arg \min }\limits_{\boldsymbol{W}(k)} &\left( {\sum\limits_{k = 0}^{N - 1} {E\{ {{{\left\| {\boldsymbol{H}(k)\boldsymbol{W}(k)\boldsymbol{s}(k) - \boldsymbol{s}(k)} \right\|}^2}} \}} } \right) \label{eqn19a}\\
&{\rm{s.t.}}\quad\boldsymbol{W}(k){\boldsymbol{P}_S}(k){\boldsymbol{W}^H}(k) = {\boldsymbol{R}_f}(k), \label{eqn19b}\\
&\quad\quad{1 \over {{N^2}}}\sum\limits_{k = 1}^N {{\boldsymbol{R}_f}(k)}  = {\boldsymbol{R}_d}. \label{eqn19c}
\end{align}
\end{subequations}

It is assumed that $E\left\{ {\boldsymbol{s}(k){\boldsymbol{s}^H}(k)} \right\} = {E_s}{\boldsymbol{I}_{{N_R}}},{\rm{ }}k = 0,1, \ldots ,N - 1$ . Therefore, problem OP2 is equivalent to
\begin{subequations}
\begin{align}
\rm{OP3:} \quad &\mathop {\arg \min }\limits_{\boldsymbol{W}(k)} \left( {\sum\limits_{k = 0}^{N - 1} {{{\left\| {\boldsymbol{H}(k)\boldsymbol{W}(k) - {\boldsymbol{I}_{{N_R}}}} \right\|}^2}} } \right) \label{eqn20a}\\
&\quad{\rm{ s.t. }}\quad{E_S} \cdot \boldsymbol{W}(k){{\boldsymbol{W}}^H}(k) = {\boldsymbol{R}_f}(k), \label{eqn20b}\\
&\quad\quad\quad{1 \over {{N^2}}}\sum\limits_{k = 1}^N {{\boldsymbol{R}_f}(k)}  = \boldsymbol{R}_d. \label{eqn20c}
\end{align}
\end{subequations}

The above optimization problem is NP-hard and cannot be solved easily \cite{byd}. Therefore, the suboptimal frequency-domain covariance matrix ${\boldsymbol{R}_f}(k)$ and the frequency domain precoding matrix $\boldsymbol{W}(k)$ are derived as follows.

\textcolor{black}{
1) \textsl{The derivation of} ${\boldsymbol{R}_f}(k)$: according to Eq. ({\ref{eqn7}}), define 
\begin{equation}
\label{eqn20to21}
{\boldsymbol{R}_f}(k) = {\alpha _k}{N^2}{\boldsymbol{R}_d}, 
\end{equation}
where $\sum\limits_{k = 0}^{N - 1} {{\alpha _k} = 1}$, and ${\alpha_k}\ge0$. Besides, if $\boldsymbol{H}(k)\boldsymbol{W}(k) = {\boldsymbol{I}_{{N_R}}}$, it has 
\begin{equation}
\left\| {{\boldsymbol{W}}(k){{\boldsymbol{W}}^H}(k) - {{\boldsymbol{H}}^ + }(k){{\left( {{{\boldsymbol{H}}^ + }(k)} \right)}^H}} \right\| = 0,
\end{equation}
where  ${{\boldsymbol{H}}^ + }(k)$ is the Pseudo-inverse of ${\boldsymbol{H}}(k)$. Thus, the optimization problem OP3 can be relaxed into
\begin{subequations}
\begin{align}
{{\rm{OP4: }}}&{\mathop {{\rm{arg min}}}\limits_{{\boldsymbol{\alpha}}} \sum\limits_{k = 1}^N {{{\left\| {\frac{{{\alpha _k}{N^2}}}{{{E_s}}}{{\boldsymbol{R}}_d} - {{\boldsymbol{H}}^ + }(k){{\left( {{{\boldsymbol{H}}^ + }(k)} \right)}^H}} \right\|}^2}} }\label{eqn21a}\\
{}&{{\rm{   s}}{\rm{.t}}{\rm{.     }}\sum\limits_{k = 1}^N {{\alpha _k}}  = 1{}}\\
{}&{{\rm{             }}\quad\quad{\alpha _k} \ge 0}
\end{align}
\end{subequations}
}
\textcolor{black}{
The optimization problem OP4 is convex \cite{byd} and can be solved by CVX tools. The suboptimal frequency-domain covariance matrix $\boldsymbol{R}_f(k)$ can be obtained according to Eq. (\ref{eqn20to21}) and the derived optimal $\boldsymbol{\alpha}$. 
}

2) {\textsl{The derivation of}} $\boldsymbol{W}(k)$: define $\boldsymbol{\Gamma}(k) = {\boldsymbol{R}_f}(k)/{E_S}$. By applying Cholesky decomposition, we can obtain
\begin{equation}
\label{eqn22}
\boldsymbol{T}(k){\boldsymbol{T}^H}(k) = \boldsymbol{\Gamma} (k),
\end{equation}
where $\boldsymbol{T}(k)$ is an invertible lower triangular matrix. Therefore, constraint (\ref{eqn20b}) can be rewritten as
\begin{equation}
\label{eqn23}
{\boldsymbol{T}^{ - 1}}(k)\boldsymbol{W}(k){\boldsymbol{W}^H}(k){\boldsymbol{T}^{ - H}}(k) = {\boldsymbol{I}_{{N_T}}}.
\end{equation}

Defining $\widetilde {\boldsymbol{W}}(k) = {\boldsymbol{T}^{ - 1}}(k)\boldsymbol{W}(k)$, the optimization problem OP3 can be split into $N$ subproblems, and the $k$th subproblem is given by
\begin{subequations}
\begin{align}
\rm{OP5:} \quad \mathop {\arg \min }\limits_{\widetilde {\boldsymbol{W}}}& \left( {{{\left\| {\boldsymbol{H}(k)\boldsymbol{T}(k)\widetilde {\boldsymbol{W}}(k) - {\boldsymbol{I}_{{N_R}}}} \right\|}^2}} \right) \label{eqn24a}\\
&{\rm{s.t.}}\quad\widetilde {\boldsymbol{W}}(k){\widetilde {\boldsymbol{W}}^H}(k) = {\boldsymbol{I}_{{N_T}}}. \label{eqn24a}
\end{align}
\end{subequations}

The optimization problem OP5 is an Orthogonal Procrustes problem (OPP), and it has a globally optimal solution, which is given by \cite{TVi}
\begin{equation}
\label{eqn25}
\widetilde {\boldsymbol{W}}(k) = \widetilde {\boldsymbol{U}}(k){\boldsymbol{I}_{{N_T} \times {N_R}}}{\widetilde {\boldsymbol{V}}^H}(k),
\end{equation}
where ${\boldsymbol{I}_{{N_T} \times {N_R}}}$ is a  diagonal matrix and each diagonal element is equal to one, $\widetilde {\boldsymbol{U}}(k)\boldsymbol{\Sigma} (k){\widetilde {\boldsymbol{V}}^H}(k)$ is the Singular Value Decomposition (SVD) of ${\boldsymbol{T}^H}(k){\boldsymbol{H}^H}(k)$ . Finally, the suboptimal precoding matrix can be derived as
\begin{equation}
\label{eqn26}
\boldsymbol{W}(k) = \boldsymbol{T}(k)\widetilde {\boldsymbol{U}}(k){\boldsymbol{I}_{{N_T} \times {N_R}}}{\widetilde {\boldsymbol{V}}^H}(k).
\end{equation}
\subsection{DFRC performance trade-off problem formulation and precoding matrix design}
Note that in the optimization problem OP2, the transmit covariance matrix strictly conforms to the optimal result of the Radar-only problem in Section III.A. Because of this, the communication performance is greatly degraded, and the interference among different information streams at the receiver is serious. Therefore, we consider a trade-off problem, and extend the traditional work in \cite{FLiu2} to the frequency-selective multipath channel, where communication performance (ISI among different data streams) and the radar performance (CRB of radar detection) are balanced.

The suboptimal precoding matrix derived in Section III.B is denoted by ${\boldsymbol{W}_d}(k)$, and the trade-off problem is formulated as \cite{FLiu2}
\begin{subequations}
\begin{align}
\rm{OP6:} \quad\mathop {\arg \min }\limits_{\boldsymbol{W}} \sum\limits_{k = 0}^{N-1} &{{\rho _1}{{\left\| {\boldsymbol{H}(k)\boldsymbol{W}(k)\boldsymbol{s}(k) - \boldsymbol{s}(k)} \right\|}^2}}   \notag\\
&+ (1 - {\rho _1}){\left\| {\boldsymbol{W}(k) - {\boldsymbol{W}_d}(k)} \right\|^2}  \label{eqn27a}\\
{\rm{s.t.}}\quad {{{E_S}} \over {{N^2}}}{\rm{tr}}&\left( {\sum\limits_{k = 0}^{N-1} {\boldsymbol{W}(k)\boldsymbol{W}^H(k)} } \right) \le {P_T}, \label{eqn27b}
\end{align}
\end{subequations}
where ${\rho _1} \in [0,1]$ is a given weighting factor that balances the radar performance and the communication performance. The optimization problem OP6 can be relaxed into $N$ simplified subproblems, and the $k$th subproblem is expressed as
\begin{subequations}
\begin{align}
\rm{OP7:} \quad&\quad\mathop {\arg \min }\limits_{\boldsymbol{W}(k)} {\rho _1}{\left\| {\boldsymbol{H}(k)\boldsymbol{W}(k) - {\boldsymbol{I}_{{N_R}}}} \right\|^2} \notag \\
&\quad+ (1 - {\rho _1}){\left\| {\boldsymbol{W}(k) - {\boldsymbol{W}_d}(k)} \right\|^2}  \label{eqn28a}\\
{\rm{s.t.}}\quad &{\rm{tr}}\left( {\boldsymbol{W}(k)\boldsymbol{W}^H(k)} \right) \le {\rm{tr}}\left( {\boldsymbol{W}_d(k)\boldsymbol{W}_d^H(k)} \right). \label{eqn28b}
\end{align}
\end{subequations}

The above optimization problem is convex and the optimal frequency-domain precoding matrix $\boldsymbol{W}(k)$  can be obtained.

\section{Joint waveform design based on CRB minimization and achievable rate maximization}
\label{s4}
In \cite{FLiu2}, \cite{FLiu3,XLiu,XWang,FLiu4} and \cite{XHu}, different receive antennas belong to different communication users, and each user is equipped with one antenna. In this paper, a single communication user MIMO scenario is considered, and all the receive antennas belong to the same user. However, there is no related integrated signal design method in the considered scenario. In this section, we propose a joint waveform design scheme based on the CRB minimization and achievable rate maximization. 
\subsection{Communication-only problem formulation}
The maximum achievable rate of the communication system is given in Eq. (\ref{eqn13}). Under the power constraint, the communication-only optimization problem can be formulated as
\begin{subequations}
\begin{align}
&{\rm{OP8:}}\mathop {\arg \max }\limits_{{\boldsymbol{W}}(k),{\boldsymbol{P}_S}(k)}  \bigg\{ \sum\limits_{k = 0}^{N - 1} {{\Delta f} \over B} \notag \\
&\cdot{{{\log }_2}\left( {\det \left( {\boldsymbol{I} + {{\boldsymbol{H}(k)\boldsymbol{W}(k){\boldsymbol{P}_S}(k){\boldsymbol{W}^H}(k){\boldsymbol{H}^H}(k)} \over {\sigma _{\boldsymbol{Z}}^2}}} \right)} \right)}\bigg\} \label{eqn29a} \\
&\quad\quad\quad{\rm{s.t.}}\quad\quad {1 \over {{N^2}}}\sum\limits_{k = 0}^{N - 1} {{\rm{tr}}({\boldsymbol{W}}(k){\boldsymbol{P}_S}(k){\boldsymbol{W}^H}(k))}  = P_T. \label{eqn29b}
\end{align}
\end{subequations}

The above optimization problem OP8 can be solved through the traditional water-filling algorithm, and according to Eq. (\ref{eqn5}), the optimal frequency-domain covariance matrix is
\begin{equation}
\label{eqn30b}
\boldsymbol{R}_{f,com}(k)= \boldsymbol{W}_{com}(k){\boldsymbol{P}_{S,com}}(k){\boldsymbol{W}^H_{com}}(k), 
\end{equation}
where $\boldsymbol{W}_{com}(k)$ is the optimal precoding matrix and ${\boldsymbol{P}_{S,com}}(k)$ is the corresponding power allocation matrix. According to the traditional water-filling algorithm, $\boldsymbol{W}_{com}(k)$ is given by
\begin{equation}
\label{eqn30to31}
\boldsymbol{W}_{com}(k) = \boldsymbol{V}_{com}(k),
\end{equation}
where ${\boldsymbol{U}_{com}}(k)\boldsymbol{\Sigma}_{com} (k){{\boldsymbol{V}^H_{com}}}(k)$ is the SVD of $\boldsymbol{H}(k)$.
\subsection{Covariance matrix design for DFRC performance trade-off}
In this subsection, we balance the communication performance (achievable rate) and the radar performance (CRB of radar detection), and propose an algorithm based on the water-filling algorithm and sequential quadratic programming (SQP) method, which is referred to as the WF-SQP algorithm, to derive the covariance matrix of the transmit signal.

We divide the time-domain covariance matrix $\boldsymbol{R}$ into two parts to achieve the performance trade-off, that is,
\begin{equation}
\label{eqn31}
\boldsymbol{R} = {\boldsymbol{R}_1} + {\boldsymbol{R}_2},
\end{equation}
where ${\boldsymbol{R}_1}$ and ${\boldsymbol{R}_2}$ are all nonnegative definite hermite matrix. The corresponding power are defined as
\begin{equation}
\label{eqn32}
\begin{cases}
{P_1} ={\rm{tr}}(\boldsymbol{R}_1) \\
{P_2} = {\rm{tr}}(\boldsymbol{R}_2) \\
{P_T} = {P_1} + {P_2}.
\end{cases}
\end{equation}

Define ${\rho _2} = {{{P_{1}}} \mathord{\left/
{\vphantom {{{P_{1}}} {{P_2}}}} \right.
 \kern-\nulldelimiterspace} {{P_T}}}$  as the trade-off factor in this waveform design scheme. Note that the total power ${P_T}$  is divided into two parts ($P_1$ and $P_2$) for optimization, and both parts are employed for the radar system and the communication system. The derivation of the frequency domain covariance matrix is as follows.

\textbf{\textsl{Step 1:}} Substitute $P_1$  for $P_T$ in Eq. (\ref{eqn29b}), and the corresponding optimal frequency-domain covariance matrix $\boldsymbol{R}_{f,1}(k)$ can be derived based on the water-filling algorithm.

\textbf{\textsl{Step 2:}} Consider the performance trade-off, and set ${\boldsymbol{R}_2} = (1 - {\rho _2}){\boldsymbol{R}_d}$, According to Eq. (\ref{eqn20to21}), the corresponding frequency-domain covariance matrix at each subcarrier is set as  ${\boldsymbol{R}_{f,{2}}}(k) = {\omega _k}{N^2}{\boldsymbol{R}_{2}}$, where ${{\omega _k}}\ge0$ is the weighting coefficient for each subcarrier and $\sum\limits_{k = 0}^{N - 1} {{\omega _k} = 1} $.

\textbf{\textsl{Step 3:}} Based on the above two steps, the optimization problem to maximize the achievable rate is given by
\begin{subequations}
\begin{align}
&  {\rm{OP9:}}\quad \mathop {\arg \max }\limits_{\boldsymbol{\omega}} \sum\limits_{k = 0}^{N - 1} {{{\Delta f} \over B}}  \notag\\
& \cdot {{{\log }_2}\left( {\det \left( {\boldsymbol{I} + {{\boldsymbol{H}(k)\left( {{\boldsymbol{R}_{f,{\rm{1}}}}(k) + {\omega _k}{N^2}{\boldsymbol{R}_{\rm{2}}}} \right){\boldsymbol{H}^H}(k)} \over {\sigma^2_{\boldsymbol{Z}}}}} \right)} \right)} \label{eqn33a}\\
&\quad\quad\quad\quad\quad\quad{\rm{s.t.}}\quad\quad \sum\limits_{k = 0}^{N - 1} {{\omega _k} = 1}, \label{eqn33b} \\
&\quad\quad\quad\quad\quad\quad{\omega _k} \ge 0,k = 0,1,2, \ldots ,N - 1, \label{eqn33c}
\end{align}
\end{subequations}
where $\boldsymbol{\omega}  = {[{\omega _0},{\omega _1}, \ldots ,{\omega _{N - 1}}]^T}$ , and the optimal $\boldsymbol{\omega}$  can be obtained by the following SQP method.

Define the optimization objective function as
\begin{align}
& \quad\quad f(\boldsymbol{\omega})= -\sum\limits_{k = 0}^{N - 1} {{{\Delta f} \over B}} \notag \\
& \cdot{{{\log }_2}\left( {\det \left( {\boldsymbol{I} + {{\boldsymbol{H}(k)\left( {{\boldsymbol{R}_{f,{\rm{1}}}}(k) + {\omega _k}{N^2}{\boldsymbol{R}_{\rm{2}}}} \right){\boldsymbol{H}^H}(k)} \over {\sigma^2_{\boldsymbol{Z}}}}} \right)} \right)}, \label{eqn34} 
\end{align}
and the constraints are
\begin{subequations}
\begin{align}
&\quad\quad\upsilon (\boldsymbol{\omega} ) = \sum\limits_{k = 0}^{N - 1} {{\omega _k} - 1}  = \boldsymbol{1}\boldsymbol{\omega} - 1, \label{eqn34plusa} \\
&\quad\quad{g_k}(\boldsymbol{\omega} ) =  - {\omega _k} =  - {\boldsymbol{e}_k}\boldsymbol{\omega} ,k = 0,1, \ldots ,N - 1, \label{eqn34plusb}
\end{align}
\end{subequations}
where $\boldsymbol{1} = [1,1 ,\cdots, 1]$, and ${\boldsymbol{e}_k} = [0,0, \cdots, \underbrace 1_{k {\rm{th}}} ,\cdots ,0]$ is the unit vector. The optimization problem OP9 can be rewritten by
\begin{subequations}
\begin{align}
& {\rm{OP10:}}\quad\mathop {\arg \min }\limits_\omega  f(\boldsymbol{\omega} )  \label{eqn35a}\\
& \quad\quad{\rm{s.t.}}\quad\upsilon (\boldsymbol{\omega} ) = 0, \label{eqn35b}\\
& \quad\quad\quad\quad\,{g_k}(\boldsymbol{\omega} ) \le 0,k = 0,1, \ldots ,N - 1. \label{eqn35c}
\end{align}\
\end{subequations}

The nonlinear function $f(\boldsymbol{\omega} )$ can be transformed into quadratic functions at the iteration point ${\boldsymbol{\omega} _{(m)}}$  through Taylor expansion, and the objective function is given by
\begin{align}
\quad {1 \over 2}{[\boldsymbol{\omega}  - {\boldsymbol{\omega} _{(m)}}]^T}{\nabla ^2}f({\boldsymbol{\omega} _{(m)}})[\boldsymbol{\omega}  - {\boldsymbol{\omega} _{(m)}}] \notag\\
+ \nabla f{({\boldsymbol{\omega} _{(m)}})^T}[\boldsymbol{\omega}  - {\boldsymbol{\omega} _{(m)}}]. \label{eqn36}
\end{align}

Define $\boldsymbol{\eta}  = \boldsymbol{\omega}  - {\boldsymbol{\omega} _{(m)}}$, and OP10 is equivalent to:
\begin{subequations}
\begin{align}
{\rm{OP11:}}&\quad\mathop {\arg \min } \limits_{\boldsymbol{\eta}}  {1 \over 2}{\boldsymbol{\eta} ^T}{\nabla ^2}f({\boldsymbol{\omega} _{(m)}})\boldsymbol{\eta}  + \nabla f{({\boldsymbol{\omega} _{(m)}})^T}\boldsymbol{\eta}  \label{eqn37a}\\
&\quad{\rm{s.t.}} \quad \quad 1\boldsymbol{\eta}  = 1 - 1{\boldsymbol{\omega} _{(m)}},\label{eqn37b}\\
& \quad - {\boldsymbol{e}_k}\boldsymbol{\eta}  \le {\boldsymbol{e}_k}{\boldsymbol{\omega} _{(m)}},k = 0,1, \ldots ,N - 1. \label{eqn37c}
\end{align}
\end{subequations}

\textcolor{black}{
The optimization problem OP11 is a standard Quadratic Programming (QP) problem with respect to $\boldsymbol{\eta}$, and it can be solved by using Lagrange multiplier method.}
The iterative steps of the SQP algorithm are summarized in Algorithm I.

\begin{table}[h]
\centering
\begin{tabular}{l}
\toprule
Algorithm I: Algorithm for solving OP9\\
\midrule
\textbf{Input:} convergence precision $\epsilon$, $\boldsymbol{H}$, $\boldsymbol{R}_{f,1}(k)$, $\boldsymbol{R}_{2}$, $\sigma^{2}_{\boldsymbol{Z}}$, $N$.\\
\textbf{Output:} ${\boldsymbol{\omega} _{(m)}}$.\\
1.	Initialize ${\boldsymbol{\omega} _{(0)}}={1 \over N}\boldsymbol{1}$, $\left\| {{\boldsymbol{\omega} _{(1)}} - {\boldsymbol{\omega}  _{(0)}}} \right\| > \varepsilon $, $m=0$.\\
\textbf{while} $\left\| {{\boldsymbol{\omega} _{(1)}} - {\boldsymbol{\omega}  _{(0)}}} \right\| > \varepsilon $\\
2.	Transform the original problem OP9 to a QP problem \\
\quad OP11 at the iteration point  ${\boldsymbol{\omega} _{(m)}}$.\\
3.	Compute ${\nabla ^2}f({\boldsymbol{\omega} _{(m)}})$ through Davidon-Fletcher-Powell \\
\quad (DFP) algorithm or Broyden-Fletcher-Goldfarb-Shanno \\
\quad (BFGS) algorithm \cite{CGB}\cite{JED}. \\
4.	Solve the QP problem and obtain the optimal result ${\boldsymbol{\eta} ^*}$.\\
5.	Compute  ${\boldsymbol{\omega} _{(m+1)}}={\boldsymbol{\omega} _{(m)}}+{\boldsymbol{\eta} ^*}$.\\
6.  $m=m+1$.\\
\textbf{end while}\\
\bottomrule
\end{tabular}
\end{table}

After obtaining the optimal result $\boldsymbol{\omega}$, ${\boldsymbol{R}_{f,{\rm{2}}}}(k)$ can be derived based on ${\boldsymbol{R}_{f,{2}}}(k) = {\omega _k}{N^2}{\boldsymbol{R}_{2}}$, and the  frequency-domain covariance matrix can be derived as $\boldsymbol{R}_f(k)={\boldsymbol{R}_{f,{1}}}(k)+{\boldsymbol{R}_{f,{2}}}(k)$.
\subsection{Precoding matrix design for the DFRC performance trade-off}
For a given power $P_1$, we can obtain the optimal precoding matrix  $\boldsymbol{W}_1(k)$ based on Eq. (\ref{eqn30to31}) and the corresponding power allocation matrix ${\boldsymbol{P}_{S,1}}(k)$ through water-filling algorithm. According to Eq. (\ref{eqn30b}), $\boldsymbol{R}_{f,1}(k)$ should satisfy
\begin{equation}
\label{eqn38}
\boldsymbol{R}_{f,1}(k) =\boldsymbol{W}_1(k){\boldsymbol{P}_{S,1}}(k){\boldsymbol{W}^H_1}(k),
\end{equation}
where ${\boldsymbol{P}_{S,{\rm{1}}}}(k) = {\rm{diag}}\{{p_{{s_1},{\rm{1}}}}(k), \cdots {p_{{s_{{N_R}}},{\rm{1}}}}(k)\}$, ${p_{{s_i},{\rm{1}}}}(k)$ is the power allocated to the $i$th information stream at the $k$th subcarrier according to the water-filling algorithm, and $\sum\limits_{k = 0}^{N - 1} {{\rm{tr}}\left( {{\boldsymbol{P}_{S,{\rm{1}}}}(k)} \right)}  = {P_{\rm{1}}}$.

\textcolor{black}{
Similarly, ${\boldsymbol{R}_{f}}(k)$ should satisfy
\begin{equation}
\label{eqn40}
\boldsymbol{R}_{f}(k) =\boldsymbol{W}(k){\boldsymbol{P}_{S}}(k){\boldsymbol{W}^H}(k).
\end{equation}
For the total transmit power ${P_T}$, since ${\rho _2} = {P_1}/{P_T}$, we can set the optimal power allocation matrix ${{\boldsymbol{P}}_S}(k)$ as
\begin{equation}{{\boldsymbol{P}}_S}(k) = \frac{1}{{{\rho _2}}} \cdot {{\boldsymbol{P}}_{S,1}}(k).
\end{equation}
Define ${\boldsymbol{P}}^\circ (k) = {({{\boldsymbol{P}}_S}(k))^{1/2}}$, and ${{\boldsymbol{R}}_f}(k)$ can be derived as
\begin{equation}
\label{eqn42}
{{\boldsymbol{R}}_f}(k) = {\boldsymbol{W}}(k){\boldsymbol{P}}^\circ (k){\left( {{\boldsymbol{W}}(k){\boldsymbol{P}}^\circ (k)} \right)^H}.
\end{equation}
}
\textcolor{black}{
The sizes of ${{\boldsymbol{R}}_f}(k)$, ${\boldsymbol{W}}(k)$, ${{\boldsymbol{P}}^ \circ }(k)$ are $N_T \times N_T$, $N_T \times N_R$, and $N_R \times N_R$, respectively. Note that ${{N}_{R}<{N}_{T}}$ is assumed in this paper. Therefore, through eigenvalue decomposition, we can derive the solution of the precoding matrix ${\boldsymbol{W}}(k)$ as \cite{GStra}
\begin{equation}
\label{eqn47}
{\boldsymbol{W}}(k) = {\boldsymbol{Q}}(k){\left( {{\boldsymbol{D}}(k)} \right)^{1/2}}{\boldsymbol{I}_{{N_T} \times {N_R}}}{\left( {{\boldsymbol{P}^ \circ }(k)} \right)^{ - 1}},
\end{equation}
where ${\boldsymbol{Q}}(k){\boldsymbol{D}}(k)\boldsymbol{Q}^{ - 1}(k) = {\boldsymbol{R}_{f}}(k)$ is the eigenvalue decomposition of  ${\boldsymbol{R}_{f}}(k)$.
}


In summary, the waveform design scheme based on the WF-SQP algorithm is provided in Algorithm II.
\begin{table}[h]
\centering
\begin{tabular}{l}
\toprule
Algorithm II: The waveform design scheme based on the \\
CRB minimization and achievable rate maximization \\
\midrule
\textbf{Input:} convergence precision $\epsilon$, $\boldsymbol{H}$, $N$, $P_T$, $\theta$, $\rho_{2}$.\\
\textbf{Output:} $\boldsymbol{W}(k),k = 0,1, \ldots ,N - 1$.\\
1.  Solve the Radar-only problem OP1 and obtain $\boldsymbol{R}_d$.\\
2.	For a given $\rho_{2}$, calculate $P_1$ and solve the Communication \\
\quad -only problem OP8 under the power constraint $P_1$. Obtain\\
 \quad the optimal ${\boldsymbol{R}_{f,{\rm{1}}}}(k)$, $\boldsymbol{W}_1(k)$, and ${\boldsymbol{P}_{S,1}}(k)$ by using the water\\
\quad -filling algorithm.\\
3.	Set \!${\boldsymbol{R}_2} = (1 - {\rho _2}){\boldsymbol{R}_d}$. \!Based \!on \!${\boldsymbol{R}_2}$ and \!${\boldsymbol{R}_{f,{\rm{1}}}}(k)$, solve the \\
\quad  problem OP9 through SQP method and obtain $\boldsymbol{\omega}_k$. \\
\textcolor{black}{4. According to ${\boldsymbol{R}_{f,{2}}}(k) = {\omega _k}{N^2}{\boldsymbol{R}_{2}}$, calculate}\\
\textcolor{black}{\quad ${{\boldsymbol{R}}_f}(k) = {{\boldsymbol{R}}_{f,1}}(k) + {{\boldsymbol{R}}_{f,2}}(k)$.}\\
5.  \textcolor{black}{Derive the precoding matrix $\boldsymbol{W}(k)$ according to Eq. (\ref{eqn47}).}\\
\bottomrule
\end{tabular}
\end{table}

It is noteworthy that PAPR is one of the most important factors in both communication and radar systems, and traditional PAPR reduction schemes \cite{GRen}\cite{IBai} can be employed in the proposed DFRC system to ensure the effectiveness of radar detection and communication service.
\subsection{Complexity analysis}
The idea of SQP algorithm is to solve quadratic programming problems through multiple iterations. In the Lagrange method employed in the proposed scheme, we need to solve a $\left( {N + M} \right) \times \left( {N + M} \right)$ linear equation system, where $M$ is the number of constraints. The complexity of solving it using Gaussian elimination method is $O({(N + M)^3})$. In this paper, it is assumed that $M=2$ and the number of convergency iterations of the SQP algorithm is $K_{\rm{SQP}}$ , and the complexity of the SQP algorithm is $O({K_{\rm{SQP}}}{(N + 2)^3})$.

For the optimization problems considered in this paper, we can also employ other nonlinear optimization algorithms, such as the Interior Point Method (IPM). The complexity analysis of IPM is as follows:

In each iteration of IPM, solving the Newton equation dominates the time complexity. The complexity of solving the Newton equation is $O(M{N^3} + {M^2}{N^2} + {M^3} + {N^3})$\cite{GDP}. It is assumed the number of convergency iterations is $K_{\rm{IPM}}$, and $M=2$. Therefore, the complexity of IPM is $O({K_{{\rm{IPM}}}}(3{N^3} + 4{N^2}))$.

Increasing the number of OFDM subcarriers $N$ will mainly benefit the communication system, because it will enhance system anti-interference ability and improve the frequency efficiency. However, it should be noted that increasing the number of subcarriers can also increase the complexity and power consumption of the DFRC system, and may have a certain impact on transmission delay. Therefore, the number of subcarriers needs to be flexibly adjusted according to the actual scenario.

In this paper, the optimal point derived by IPM and SQP are consistent. Because SQP has a lower complexity, we employ the SQP algorithm in the proposed scheme.
\section{Simulation results}
\label{s5}

In this section, we verify the proposed waveform design schemes through simulations. The total transmit power $P_T$ is normalized to be 1, and each entry of the frequency-domain channel matrix ${\boldsymbol{H}(k)}$ is subject to standard complex Gaussian distribution. It is noteworthy that more standardized channel model will be adopted in our future work to verify the performance of the proposed scheme. The DFRC BS is equipped with $N_T=20$ antennas. The number of receive antennas at the communication user is $N_R=10$. The data streams are modulated by quadrature phase shift keying (QPSK). The number of the OFDM subcarriers is $N=16$. The interested detection target is located at the angle of ${0^ \circ }$. In the figures, the traditional scheme in \cite{FLiu2} labelled with `Traditional scheme' \textcolor{black}{and the JCAS scheme in \cite{JCAS} labelled with `JCAS scheme'} are provided as the benchmark schemes, the designs with the strict CRB minimization constraint (mentioned in Section \ref{s3} B) and the performance trade-off are labelled with `Strict' and `Trade-off', respectively, and we use `ISI-min' (inter-stream interference minimization), and `AR-max' (achievable rate maximization) to distinguish the proposed two schemes. \textcolor{black}{In the simulation, the trade-off factor $\rho$ corresponds to $\rho_1$ in the 'ISI-min' scheme and $\rho_2$ in the 'AR-max' scheme, respectively}. Note that the precoding matrix design corresponding to `AR-max' scheme in Eq. (\ref{eqn47}) is an approximate solution, thus the derived achievable rate using the proposed precoding design will be lower than the ideal optimal result of OP9, and these two cases are labelled as `Actual' and `Ideal', respectively.

\begin{figure}[tbp]
\centering
\includegraphics [width=0.9\linewidth]{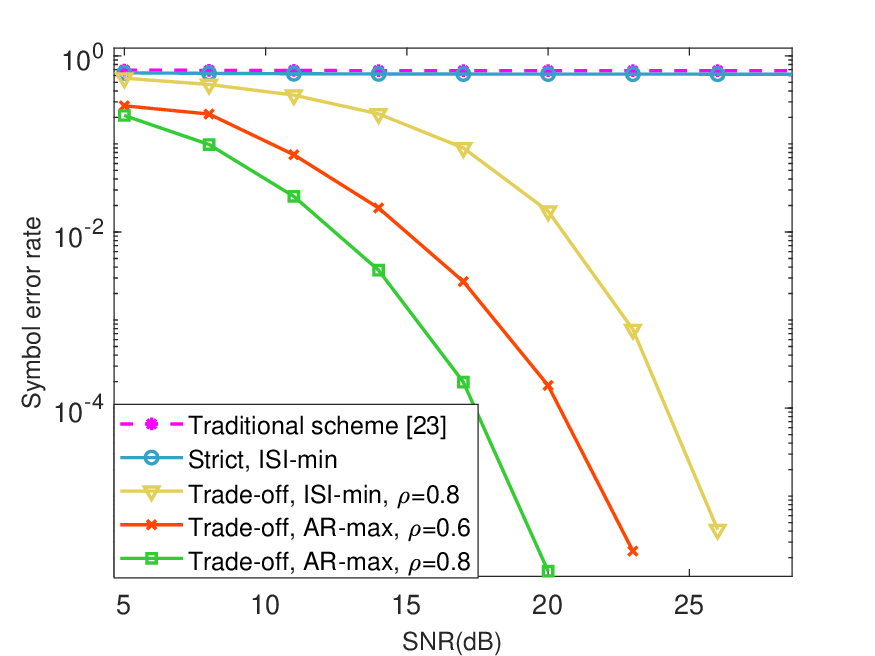}
\caption{SER comparison of different schemes with different SNR.}
\label{Fig1}
\end{figure}

The SER of the proposed waveform design schemes are plotted in Fig. \ref{Fig1}. Note that the frequency-selective multipath fading channel is considered in this paper, the scheme in \cite{FLiu2} labelled by `Traditional scheme' cannot be directly applied to achieve a satisfied result. It can be seen from  Fig. \ref{Fig1} that the `Strict' design suffers great communication performance loss under the strict equality constraint (\ref{eqn19a})(\ref{eqn19b}), and the communication performance can be improved by introducing the trade-off factor. Fig. \ref{Fig1} also shows the performance of the proposed `Trade-off' design scheme based on the CRB and ISI minimization, which is an extension of the work in \cite{FLiu2} to the frequency-selective multipath fading environment. It can be also seen from Fig. \ref{Fig1} that the second proposed waveform design scheme based on the CRB minimization and the achievable rate maximization achieves the best SER performance.

\begin{figure}[tbp]
\centering
\includegraphics [width=0.9\linewidth]{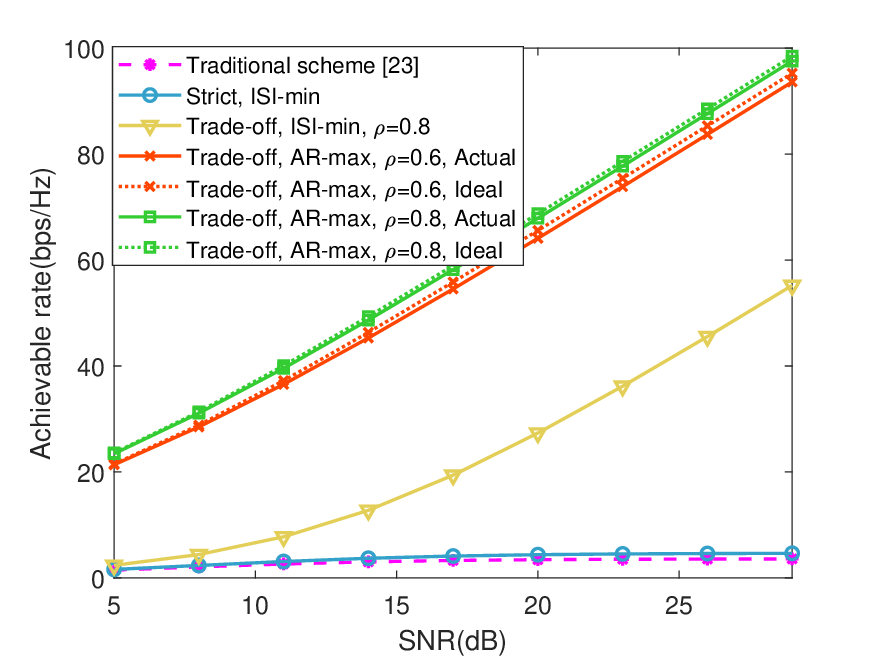}
\caption{Achievable rate comparison of different schemes with different SNR.}
\label{Fig2}
\end{figure}

\begin{figure}[tbp]
\centering
\includegraphics [width=0.9\linewidth]{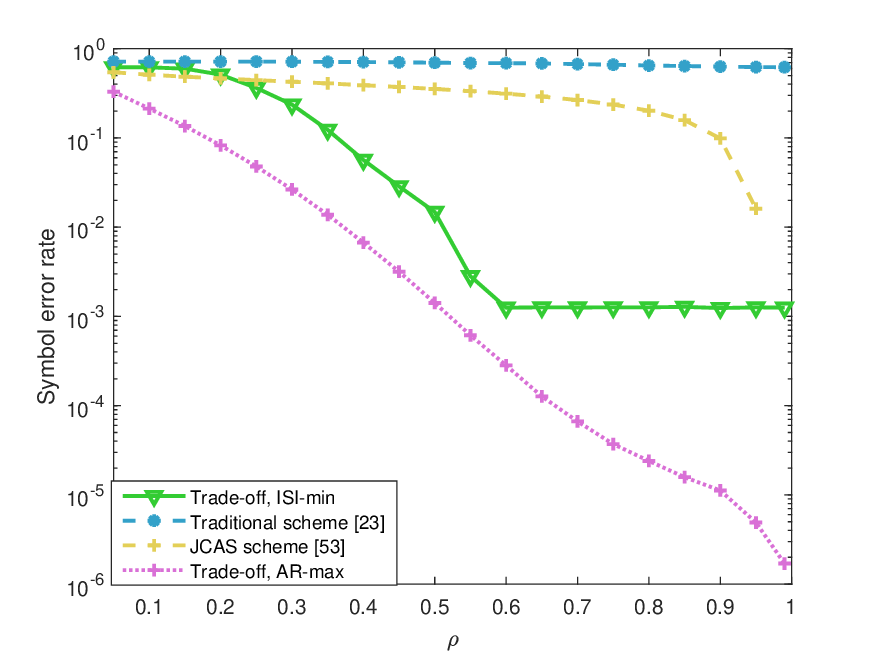}
\caption{SER comparison of different schemes with different trade-off factors and SNR=20 dB.}
\label{Fig3}
\end{figure}

\begin{figure}[tbp]
\centering
\includegraphics [width=0.9\linewidth]{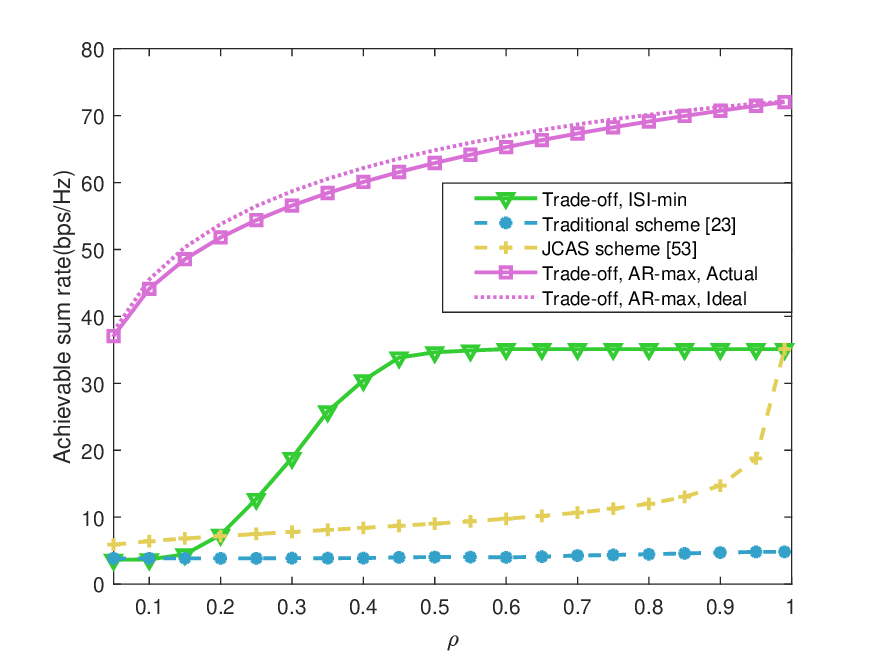}
\caption{Achievable rate comparison of different schemes with different trade-off factors and SNR=20 dB.}
\label{Fig4}
\end{figure}

\begin{figure}[tbp]
\centering
\includegraphics [width=0.9\linewidth]{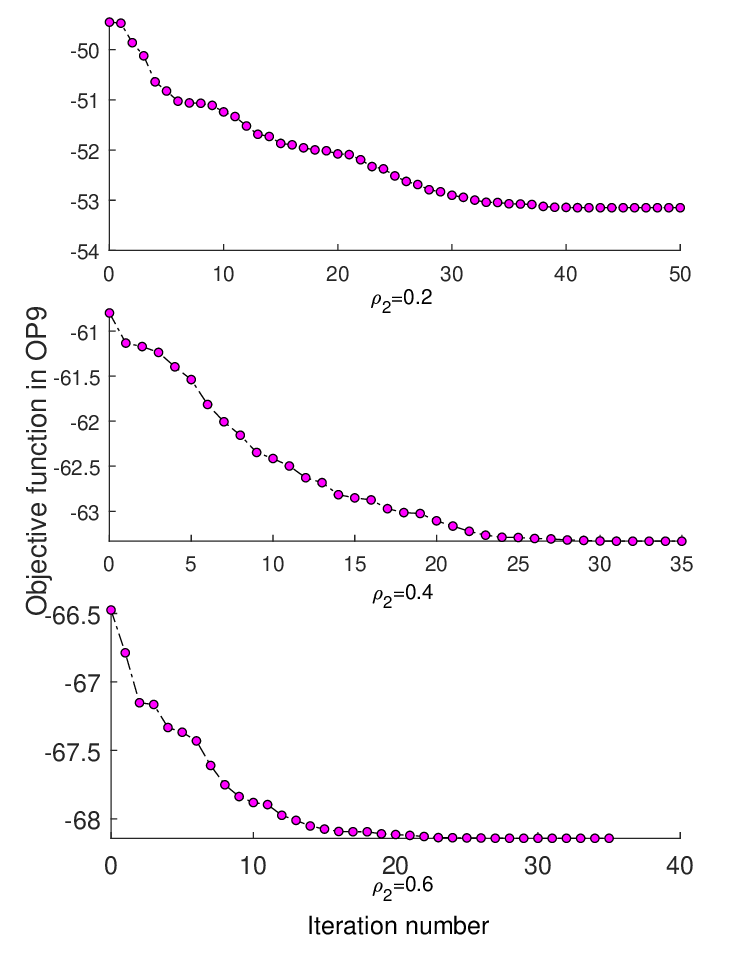}
\caption{Convergence performance verification of the proposed WF-SQP algorithm.}
\label{fmin_p_diedai}
\end{figure}

\begin{figure}[tbp]
\centering
\includegraphics [width=0.9\linewidth]{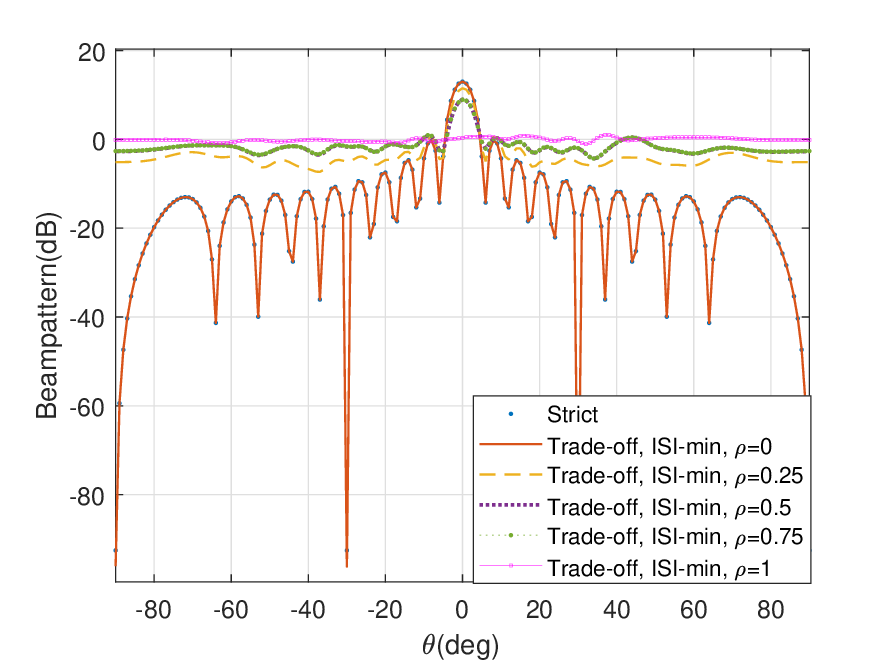}
\caption{Radar beam pattern obtained by the `ISI-min' scheme with different trade-off factors.}
\label{Fig5}
\end{figure}

\begin{figure}[tbp]
\centering
\includegraphics [width=0.9\linewidth]{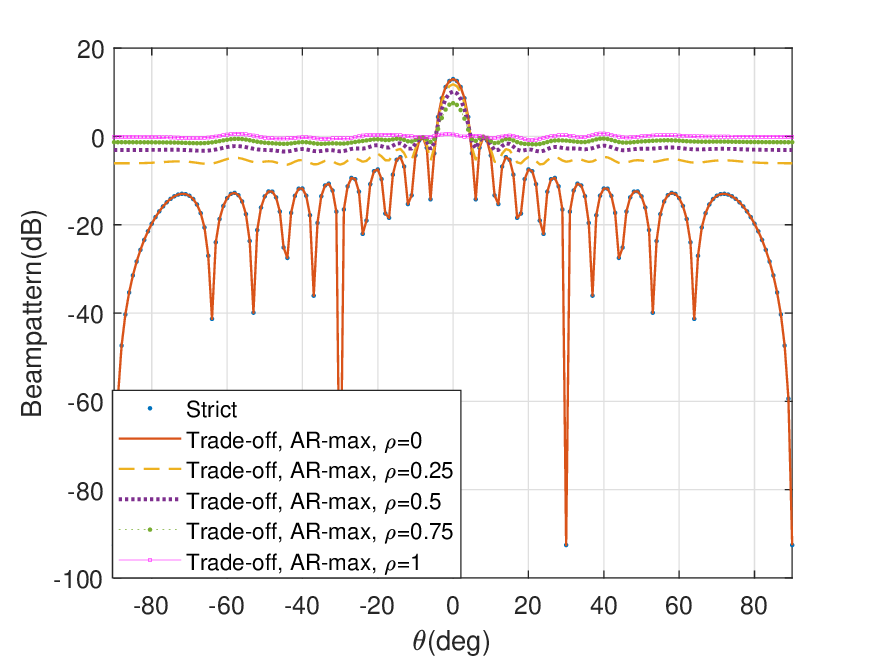}
\caption{Radar beam pattern obtained by the `AR-max' scheme with different trade-off factors.}
\label{Fig6}
\end{figure}

Fig. \ref{Fig2} shows the achievable rate of the proposed waveform design schemes. It can be seen from Fig. \ref{Fig2} that the proposed waveform design scheme based on the achievable rate maximization outperforms the scheme based on ISI minimization. Furthermore, it is noteworthy that by employing the proposed precoding matrix design in Eq. (\ref{eqn47}), the achievable rate only experiences a slight performance loss compared to the ideal optimal result of OP9.

The SER versus the trade-off factor is shown in Fig. \ref{Fig3}, where SNR is $20 \,\rm{dB}$. It can be seen from Fig. \ref{Fig3} that the SERs of the proposed two waveform design schemes decrease with the increase of the trade-off factor. \textcolor{black}{Since the JCAS scheme in \cite{JCAS} is based on the multi-carrier modulation and the scheme in [23] is only suitable for the flat fading channel, the SER of the JCAS scheme is lower than that of the scheme in [23], but higher than that of the proposed two schemes.} In the first waveform design scheme (ISI-min), when the trade-off factor is greater than 0.6, there is an error floor. That is because when the first term in Eq. (\ref{eqn28a}) is close to 0, there is a communication performance bound of this scheme. For the second waveform design scheme (AR-max), there is no error floor, which proves the superiority of the proposed `AR-max' scheme.

Fig. \ref{Fig4} shows achievable rate versus the trade-off factor, and SNR is also equal to 20 dB. The achievable rates of the two waveform design schemes get improved with the increase of trade-off factor. \textcolor{black}{The achievable rate of the JCAS scheme is higher than that of the scheme in [23], but lower than the proposed two schemes.} Moreover, the communication performance of the `AR-max' scheme continuously improves with the increase of trade-off factor, while the `ISI-min' based scheme reaches its performance bound when the trade-off factor is greater than 0.6, which is similar to the results in Fig. \ref{Fig3}. Meanwhile, the stability of the second proposed waveform design scheme (AR-max) is also verified in Fig. \ref{Fig4}, and the gap between the actual result and the ideal optimal result in OP9 is very small.

It is noteworthy that the `AR-max' scheme and `ISI-min' scheme are fundamentally different. The aim of `ISI-min' scheme is to minimize the interference among the signals of different receive antennas, and the signals of different receive antennas do not need to be jointly processed. The aim of the `AR-max' scheme is to maximize the achievable sum rate, and the receiver jointly processes the signals of different receive antennas, which achieves better performance than the `ISI-min' scheme. Therefore, the SER and achievable rate performances of the `AR-max' scheme
 are better than those of the `ISI-min' scheme.

Fig. \ref{fmin_p_diedai} shows the variation of the objective function in OP9 with the iteration number of the WF-SQP algorithm. The convergence performance of different trade-off factors are compared in Fig. \ref{fmin_p_diedai}. The convergence precision $\epsilon$ is set to be $10^{-6}$, and SNR is $20 \,\rm{dB}$. It can be seen from Fig. \ref{fmin_p_diedai} that for each trade-off factor, the objective function decreases with the increase of iteration number. Furthermore, the termination conditions are satisfied within 40 iterations. Therefore, the convergence performance of the proposed WF-SQP algorithm can achieve a good performance.

The resultant radar beampatterns obtained by the two proposed schemes with different trade-off factors are provided in Fig. \ref{Fig5} and Fig. \ref{Fig6}, respectively. The `Strict' beam pattern corresponds to the optimal result of OP1. It can be seen from Fig. \ref{Fig5} and Fig. \ref{Fig6} that when the trade-off factor equals 0, the proposed two performance trade-off waveform design schemes obtain the same beam pattern as the `Strict' beam pattern. Meanwhile, with the increase of trade-off factor, the mismatch between the obtained beam pattern and the `Strict' beam pattern begins to increase, because the radar performance is sacrificed to enhance the communication performance.

\begin{figure}[t]
\centering
\begin{minipage}[t]{1\linewidth}
\centering
\includegraphics[width=1\linewidth]{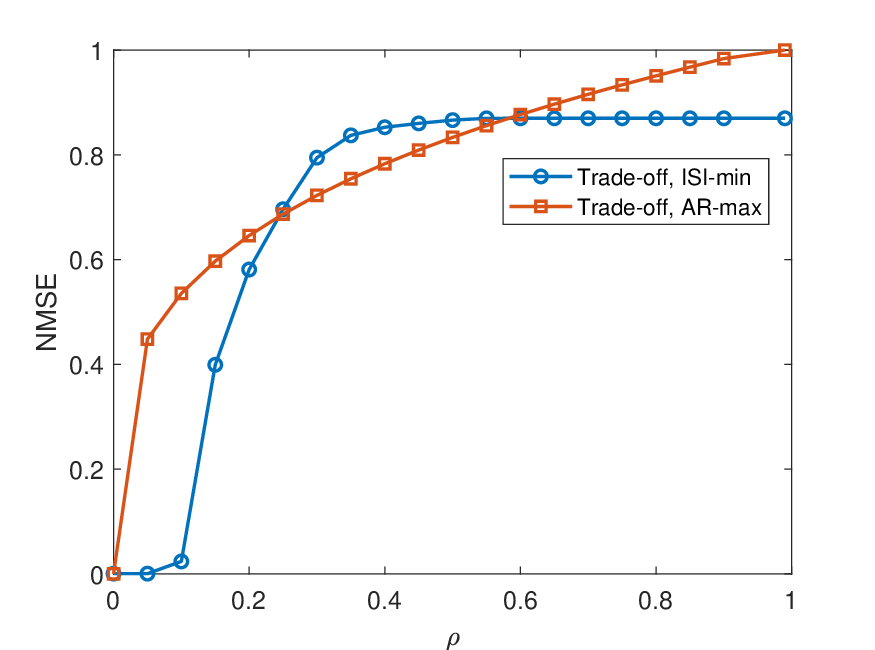}
\caption{NMSEs of radar beam pattern obtained by different schemes.}
\label{NMSE}
\end{minipage}
\begin{minipage}[t]{1\linewidth}
\centering
\includegraphics[width=1\linewidth]{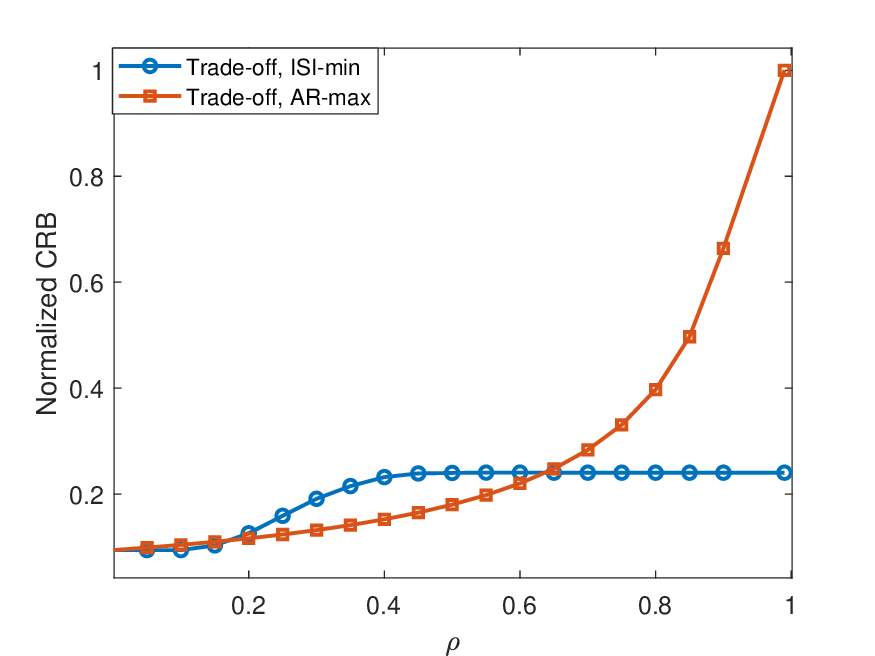}
\caption{Normalized CRBs obtained by different schemes.}
\label{CRB}
\end{minipage}
\end{figure}

We employ the normalized mean square error (NMSE) to describe the beampattern mismatch. The NMSE is defined as
\begin{equation}
\label{eqn45}
\rm{NMSE} = {{\int {{{\left\| {{B_S}(\theta ) - {B_T}(\theta )} \right\|}^2}d\theta } } \over {\int {{{\left\| {{B_S}(\theta )} \right\|}^2}d\theta } }},
\end{equation}
where ${B_S}(\theta )$ is the `Strict' beam pattern, and ${B_T}(\theta )$ is the beam pattern with `Trade-off' scheme.

The beampattern NMSEs of different schemes with different trade-off factors are plotted in Fig. \ref{NMSE}, and the normalized CRBs obtained by different schemes are provided in Fig. \ref{CRB}, which is consistent with Fig. \ref{NMSE}. It can be observed that for trade-off factor within $[0.2,0.6]$, the NMSE of the `AR-max' scheme is lower than that of the `ISI-min' scheme, which means a lower CRB for the target detection. From Fig. \ref{Fig3} and Fig. \ref{Fig4}, when the trade-off factor is less than $0.2$, the communication performance is not acceptable (with SER$\ge10\%$); when the trade-off factor is greater than $0.6$, the communication performance of the `ISI-min' scheme keeps the same, which means the communication performance cannot be improved with the increase of the trade-off factor. Therefore, the trade-off factor works in the range of $[0.2,0.6]$ for the `ISI-min' scheme. Besides, it can be seen from Fig. \ref{NMSE} and Fig. \ref{CRB} that the `AR-max' scheme has better radar performance when the trade-off factor is within $[0.2,0.6]$. Therefore, the proposed the `AR-max' scheme can achieve a better performance trade-off between the radar system and the communication system.

\begin{figure}[tbp]
\centering
\includegraphics [width=1\linewidth]{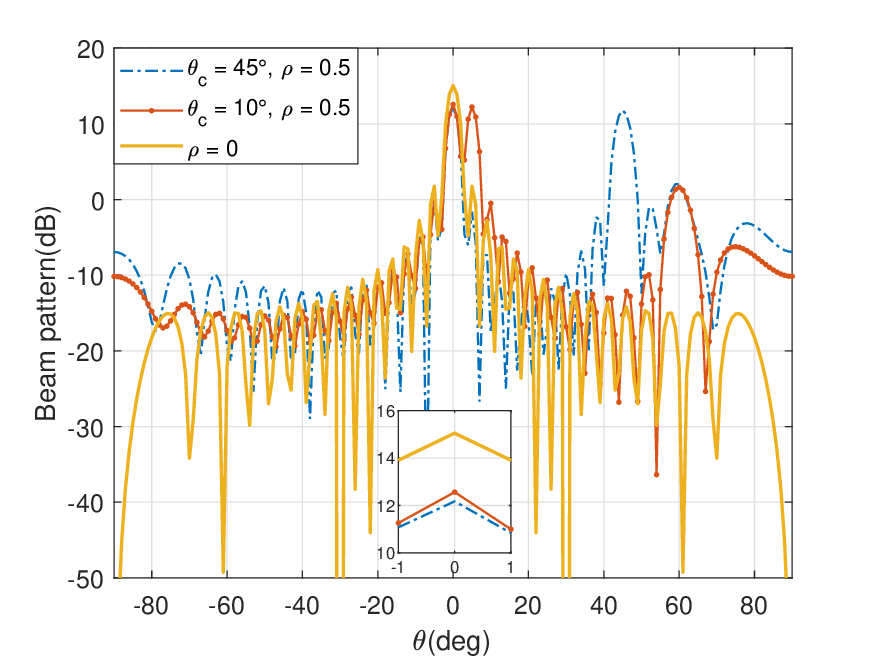}
\caption{Radar beam pattern with different $\theta_{c}$.}
\label{Figa}
\end{figure}

\begin{figure}[tbp]
\centering
\includegraphics [width=1\linewidth]{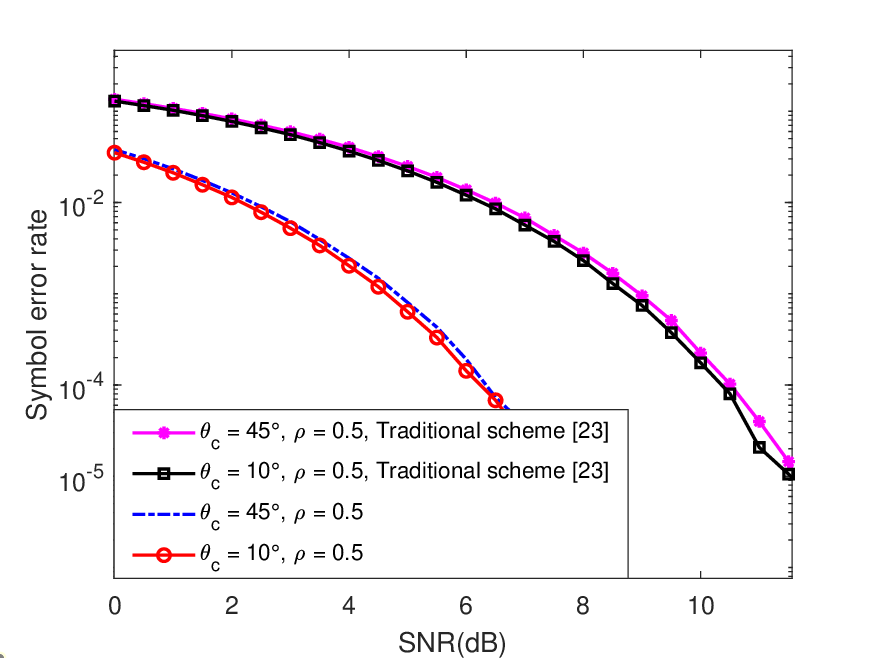}
\caption{SER with different $\theta_{c}$.}
\label{Figb}
\end{figure}

Fig. \ref{Figa} and Fig. \ref{Figb} show the impact of the angular spacing between the detection target and the communication user on the radar and communication performance.  It is assumed that the communication user is located at the angle of $\theta_{c}$ , and the detection target is located at ${0^ \circ }$. By employing the ‘AR-max’ scheme, the transmit beam pattern with different $\theta_{c}$ is shown in Fig. \ref{Figa}. It can be seen from Fig. \ref{Figa} that when $\rho=0$, that is, the communication performance is not considered, the tramsmit beam radiates highest power to the detection target. When $\rho=0.5$, that is, the radar detection performance and the communication performance are jointly considered, the transmit beam pattern radiates lower power towards the detection target, which will decrease the radar detection performance. Besides, it can be seen from Fig. \ref{Figa} that the proposed scheme is robust to the angular spacing between the detection target and the communication user. Fig. \ref{Figb} shows the SER with different $\theta_{c}$. The SER performance of the traditional scheme using single carrier modulation \cite{FLiu2} is also provided in Fig. \ref{Figb} for comparison. It can be seen from Fig. \ref{Figb} that the proposed waveform design scheme has a better SER performance than the traditional scheme. Fig. \ref{Figb} also shows that the proposed waveform design scheme is robust to the angular spacing between the detection target and the communication user.

\vspace{-0.1in}
\section{Conclusion}
\label{s6}
Considering the communication and radar detection performance trade-off, this paper proposed two waveform design schemes for the DFRC system based on MIMO and OFDM technologies in the frequency-selective multipath fading environment. ISI and achievable rate were employed as the performance metrics in the two proposed design schemes, respectively. In the CRB and ISI minimization based waveform design scheme, which can be viewed as an extension of the traditional single carrier based DFRC waveform to the frequency-selective multipath fading environment, we first solved the radar-strict problem and then allowed a tolerable deviation between the actual precoding design and the radar-strict design to achieve the performance trade-off between the radar system and the communication system. In the CRB minimization and achievable rate maximization based waveform design scheme, we proposed the WF-SQP algorithm to solve the optimization problem. Simulation results showed the superiority of the proposed waveform design schemes.

\vspace{-0.1in}

\bibliographystyle{IEEE}

\end{document}